\documentclass[%
 reprint,
 amsmath,amssymb,
 aps,
]{revtex4-2}

\usepackage[usenames]{color}
\usepackage{colortbl}
\usepackage{graphicx}% Include figure files
\usepackage{dcolumn}% Align table columns on decimal point
\usepackage{bm}% bold math
\begin{document}

\preprint{APS/123-QED}

\title{Exceptional points of arbitrary high orders induced by non-Markovian dynamics
}% Force line breaks with \\

\author{Timofey T. Sergeev}
\affiliation{Moscow Institute of Physics and Technology, 141700, 9 Institutskiy pereulok, Moscow, Russia}
\affiliation{Dukhov Research Institute of Automatics (VNIIA), 127055, 22 Sushchevskaya, Moscow, Russia}
\affiliation{Institute for Theoretical and Applied Electromagnetics, 125412, 13 Izhorskaya, Moscow, Russia}
\author{Evgeny S. Andrianov}
\affiliation{Moscow Institute of Physics and Technology, 141700, 9 Institutskiy pereulok, Moscow, Russia}
\affiliation{Dukhov Research Institute of Automatics (VNIIA), 127055, 22 Sushchevskaya, Moscow, Russia}
\affiliation{Institute for Theoretical and Applied Electromagnetics, 125412, 13 Izhorskaya, Moscow, Russia}
\author{Alexander A. Zyablovsky}
\email{zyablovskiy@mail.ru}
\affiliation{Moscow Institute of Physics and Technology, 141700, 9 Institutskiy pereulok, Moscow, Russia}
\affiliation{Dukhov Research Institute of Automatics (VNIIA), 127055, 22 Sushchevskaya, Moscow, Russia}
\affiliation{Institute for Theoretical and Applied Electromagnetics, 125412, 13 Izhorskaya, Moscow, Russia}

\date{\today}% It is always \today, today,

\begin{abstract}
Exceptional points are singularities in the spectrum of non-Hermitian systems in which several eigenvectors are linearly dependent and their eigenvalues are equal to each other. Usually it is assumed that the order of the exceptional point is limited by the number of degrees of freedom of a non-Hermitian  system. In this letter, we refute this common opinion and show that non-Markovian effects can lead to dynamics characteristic of systems with exceptional points of higher orders than the number of degrees of freedom in the system. This takes place when the energy returns from reservoir to the system such that the dynamics of the system are divided into intervals in which it describes by the product of the exponential and a polynomial function of ever-increasing order.
We demonstrate that by choosing the observation time, it is possible to observe exceptional points of arbitrary high orders.

\end{abstract}

%\keywords{ergodic dynamics, non-ergodic dynamics, discrete time crystal}%Use showkeys class option if keyword
\maketitle

%\tableofcontents
\textit{Introduction.} Exceptional points (EPs) are singularities in the spectrum of non-Hermitian systems in which several eigenvectors are linearly dependent and their eigenvalues are equal to each other \cite{miri2019exceptional,ozdemir2019parity}. The number of such linearly dependent eigenvectors determines the order of the EP \cite{moiseyev2011non}. Due to the collinearity of some of them, the eigenvectors do not form a complete basis at the EP. Therefore, to obtain a complete basis, the set of eigenvectors is complemented by adjoint vectors. A distinctive feature of non-Hermitian systems at an EP is that the temporal dynamics of oscillation amplitudes are described by the product of a polynomial function and an exponential function \cite{lin2025non}. The maximum degree in a polynomial function is determined by the order of the EP \cite{lin2025non}.

Currently, non-Hermitian systems with EPs are the subject of intense study, being of interest from both practical and fundamental perspectives \cite{miri2019exceptional,ozdemir2019parity,ding2022non,zyablovsky2014pt,doronin2019lasing,li2023exceptional}. For example, EPs significantly amplify the system's response to minor perturbations, making the system with EPs extremely useful for metrology \cite{wiersig2016sensors,chen2017exceptional,wiersig2020review,mao2024exceptional,zhang2019quantum,lai2019observation,kononchuk2022exceptional}. The EP enhancement of the sensitivity grows with the order of EP \cite{hodaei2017enhanced,wu2021high}. 

To implement the systems with EPs, it is necessary to fulfill certain relationships between relaxation rates and coupling strengths. Due to the need for fine-tuning, implementing systems with EPs is a complex task, especially for high-order. Recently, exceptional points have been considered within the non-Markovian dynamics \cite{garmon2017characteristic,garmon2021anomalous,mouloudakis2022coalescence,cheung2018emergent,khandelwal2024emergent,sergeev2023signature,lin2025non,sergeev2025non}. It has recently been demonstrated that non-Markovian effects in non-Hermitian systems can lead to the appearance of additional and higher-order EPs \cite{lin2025non,experimentalobservationEP}. In these papers, the EPs is determined by the eigenvalues of the Liouvillian superoperator \cite{lin2025non,experimentalobservationEP}. The non-Markovian effects lead to the emergence of a bosonic pseudo-mode in the reservoir. To describe the degrees of freedom of both the system and pseudo-mode the extended Liouvillian superoperator is used. Involvement in the consideration of degrees of freedom associated with the pseudo-mode leads to the emergence of additional and higher-order EPs \cite{lin2025non,experimentalobservationEP}. This approach does not require an increase in the dimensionality of the system itself, which can facilitate the creation of systems with higher-order exceptional points. For example, in the papers \cite{lin2025non,experimentalobservationEP}, third-order EP were obtained using non-Markovian effects.

In this letter, we study the system of a single harmonic oscillator interacting with a non-Markovian reservoir. We demonstrate that non-Markovian effects lead to dynamics characteristic of systems with EPs of arbitrary orders. This takes place when the energy returns from reservoir to the system such that the system dynamics ceases to be exponential. The dynamics of the system are divided into intervals, during each of which the ones are described by the product of the exponential and a polynomial function, the order of which increases with the interval number. The higher the maximum degree of a polynomial function, the higher the order of the EP during the given time interval. The deviation of the dynamics from the exponential law leads to modification of the spectrum, in particular, to the appearance of additional peaks. The number of peaks is proportional to the EP order, and the peak widths are inversely proportional to the EP order. Based on the results obtained, we demonstrate that by choosing the observation time, it is possible to observe exceptional points of arbitrary high orders.

\textit{Model.} We consider a system of an oscillator interacting with a finite-sized reservoir. The frequency of the oscillator is equal to $\omega_0$. The reservoir is described as a finite set of $N+1$ modes with equidistant distribution of frequencies that lie close to $\omega_0$ and have the form $\omega_j = \omega_0 + j \delta\omega$ ($j=-N/2,...,N/2$), where $\delta\omega$ is a step between the modes' frequencies.

To describe the system, we use the following Hamiltonian \cite{scully1997quantum}:

\begin{equation}
\begin{array}{l}
\hat H = {\omega _0}\hat a^\dag {{\hat a}} + \sum\limits_{j = -N/2}^{N/2} {{\omega_j \hat b_j^\dag {{\hat b}_j}}} + \sum\limits_{j = -N/2}^{N/2} {{g}(\hat a {{\hat b}_j^{\dag}} + \hat a^{\dag} {{\hat b}_j})}
\end{array}
\label{eq:1}
\end{equation}
Here $\hat a$ and $\hat a^{\dag}$ are the annihilation and creation operators of oscillator that obey the bosonic commutation relation $[\hat a,\hat a^{\dag}] = 1$. $\hat b_j$, $\hat b_j^{\dag}$ are the annihilation and creation operators of the modes of the reservoir that also obey the bosonic commutation relation $[\hat b_i, \hat b_j^{\dag}] = \delta_{ij}$. $g$ is the coupling strength between the oscillator and each of mode of the reservoir.

To describe the dynamics of the system, we use the Heisenberg equations for operator \cite{landau1977quantum}. After switching to the averages of the operators we obtain the following system of equations:
\begin{equation}
\frac{{d{a}}}{{dt}} =  - i{\omega _0}{a} -i\sum\limits_{j =-N/2}^{N/2} {g{b_j}}
\label{eq:2}
\end{equation}

\begin{equation}
\frac{{d{b_j}}}{{dt}} =  - i{\omega_j}{b_j} - ig{a}
\label{eq:3}
\end{equation}
where $a(t) = \langle \hat a (t) \rangle$ and $b_j(t)=\langle \hat b_j (t) \rangle$. We consider the initial condition to be $a(0)=1$ and $b_j(0) =0$.

In the case where the number of modes in the reservoir is much greater than one, we can eliminate the reservoir's degrees of freedom \cite{carmichael1999statistical,gardiner2004quantum} and obtain the non-Hermitian equation. For simplicity, we switch to slow amplitudes, which is equivalent to the case of $\omega_0=0$. Within the Born-Markov approximation \cite{carmichael1999statistical,gardiner2004quantum}, the equation takes the following form (see section A of the Supplementary Materials)
\begin{equation}
\frac{{d{a}}}{{dt}} = - \gamma {a}
\label{eq:4}
\end{equation}
where $\gamma=\sum\limits_{j=-N/2}^{N/2} {\pi g^2 \delta(\omega_j-\omega_0)} =\pi g^2/\delta\omega$ is an effective relaxation rate, obtained through elimination of the reservoir's degrees of freedom within the Born-Markov approximation \cite{carmichael1999statistical,gardiner2004quantum,scully1997quantum}. 

\begin{figure}[htbp]
\centering\includegraphics[width=0.9\linewidth]{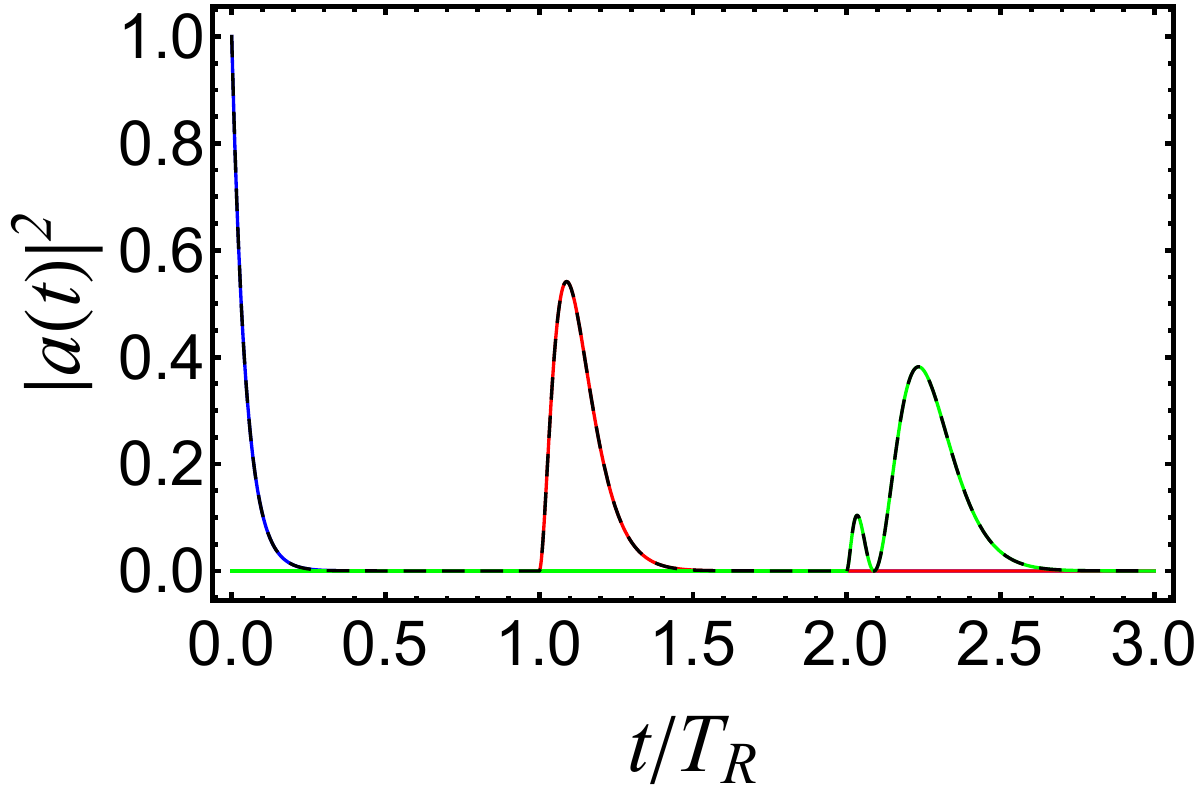}
\caption{Dynamics of $|a(t)|^2$ (the dashed black line) and the absolute values of the revival amplitudes, $a_n(t)$, determined by the Eq.~(\ref{eq:9}). The blue line shows $a_0 (t) \theta (t)$; the red line shows $a_1 (t-\frac{2\pi}{\delta\omega}) \theta (t-\frac{2\pi}{\delta\omega})$; and the green line shows $a_2 (t-\frac{4\pi}{\delta\omega}) \theta (t-\frac{4\pi}{\delta\omega})$. Here $T_R=\frac{2\pi}{\delta\omega}$ indicates the time of the first revival, $\delta\omega=0.002 \omega_0$, $\gamma \approx 0.0035 \omega_0$ and $\omega_0=1$.}
\label{fig1}
\end{figure}

The equation~(\ref{eq:4}) predicts exponential decay of the oscillator amplitude. For the complete system~(\ref{eq:2}), (\ref{eq:3}), the exponential decay is observed only at times shorter than $T_R = 2 \pi / \delta \omega$. At large times, the amplitude of the oscillator exhibits repeated revivals of oscillations over time \cite{tatarskii1987example,fleischhauer1993revivals,ferreira2021collapse, sergeev2023signature} [Figure~\ref{fig1}]. These revivals arise due to a finite size of the reservoir, which is a manifestation of non-Markovian effects in the reservoir.

\textit{Non-Markovian dynamics.} The equations~(\ref{eq:2}), (\ref{eq:3}) are linear differential equations. To calculate the system's dynamics, we find the eigenfrequencies and eigenvectors of the matrix on the right-hand side of the equations~(\ref{eq:2}), (\ref{eq:3}). The corresponding eigenfrequencies are given by the following expression (see section B of the Supplementary Materials):

\begin{equation}
\tilde\omega_k = k\delta\omega+\frac{\delta\omega}{\pi}\arctan\bigg(\frac{\gamma}{\tilde\omega_k}\bigg)
\label{eq:6}
\end{equation}

The amplitude of the oscillator is determined by the following sum (see section C of the Supplementary Materials):
\begin{equation}
a(t)=\sum\limits_{n=0}^{\infty} {\theta(t-\frac{2\pi n}{\delta\omega})a_n (t-\frac{2\pi n}{\delta\omega})}
\label{eq:7}
\end{equation}
Here $\theta(x)$ is the Heaviside step function. The $n$-th term in the sum on the right part determines the time dependence for the $n$-th revival and is given by the expression (see section C of the Supplementary Materials):
\begin{equation}
a_n(t)=\int\limits_{-\infty}^{\infty} d\omega \frac{\gamma}{\pi}\frac{e^{-i\omega t}}{\omega^2 + \gamma^2} \frac{(\omega-i\gamma)^n}{(\omega+i\gamma)^n}
\label{eq:8}
\end{equation}
The zero term corresponds to the initial stage of the system's evolution at $t < T_R$. Subsequent terms $a_n$ ($n =1, 2,...$) correspond to the sequence of revivals at moments of time $T_{R, n}= \frac{2\pi n}{\delta\omega}$.

It is seen that the integrand for the $n$-th revival contains a pole of the $(n+1)$-th order. The presence of a pole of the $(n+1)$-th order leads to the fact that the time dependence of the oscillator amplitude represents the product of a damped exponential and a polynomial function of the $n$-th order (see section C of the Supplementary Materials), namely:
\begin{equation}
a_n (t) = e^{-\gamma t} L_n^{(-1)} (2\gamma t)
\label{eq:9}
\end{equation}
Here $L_n^{(-1)} (x)$ is the generalized Laguerre polynomial of $n$-th order. Laguerre polynomials are power functions whose maximum degree coincides with their order. The difference between the evolution of the system and the exponential law is characteristic of the behavior at an exceptional point \cite{lin2025non}.

\textit{Exceptional points.} From the equation~(\ref{eq:8}) one can show that the amplitudes $a_n(t)$ satisfy the following equations (see section C of the Supplementary Materials):
\begin{equation}
\dot a_n (t)=-\gamma a_n (t) - 2\gamma \sum\limits_{k=0}^{n-1} {a_k (t)}
\label{eq:10}
\end{equation}
or, equivalently, in the matrix form:
\begin{equation}
\frac{d}{{dt}}\left( {\begin{array}{*{20}{c}}
  {{a_0}} \\ 
  {{a_1}} \\ 
  {a_2} \\ 
  {\vdots } \\
  {a_n}
\end{array}} \right) = \left( {\begin{array}{*{20}{c}}
  { - \gamma}&{ 0 }&{ 0}&{ \dots}&{0} \\ 
  { - 2\gamma }&{ -\gamma}&{ 0 }&{ \dots}&{0} \\ 
  {-2\gamma }&{-2\gamma}&{-\gamma}&{\dots}&{0} \\ 
  {\vdots }&{\vdots}&{\vdots }&{\ddots} \\
  {-2\gamma }&{-2\gamma}&{-2\gamma}&{\dots}&{-\gamma}
\end{array}} \right)\left( {\begin{array}{*{20}{c}}
  {{a_0}} \\ 
  {{a_1}} \\ 
  {a_2 } \\ 
  {\vdots } \\
  {a_n}
\end{array}} \right)
\label{eq:11}
\end{equation}
Here the initial condition is $a_0 (0)=1$ and $a_k (0)=0$, $k>0$. Note that the equation for $a_0$ coincides with the equation~(\ref{eq:4}).

One can see that this system of non-Hermitian equations has only one $(n+1)$-degenerate eigenvalue $\lambda=-\gamma$. It is seen that the first $n$ revivals are described by a matrix of size $n+1$ by $n+1$. All eigenvalues of a matrix of any size are equal to $-\gamma$. For such a matrix, there is one eigenvector and $n$ adjoint vectors. This behavior is completely consistent with an exceptional point of the $(n+1)$-th order. Thus, by observing the first $n$ revivals we detect behavior corresponding to a system with an exceptional point of the $(n+1)$-th order.

The equation~(\ref{eq:10}) implies that the dynamics of the n-th revival are determined by previous revivals, which act as an external force for it. This behavior is a consequence of the presence of memory in the system, i.e., the non-Markovian nature of the system. Thus, due to non-Markovian processes, the system under consideration exhibits behavior at different time intervals that is characteristic of behavior at exceptional points of ever-increasing order.

\textit{Method for observing exceptional points.} To observe exceptional points in the system under consideration, it is necessary to measure the amplitudes of the revivals, $a_n (t)$, separately. These amplitudes determine the amplitude of the oscillator, $a(t)$ (Eq.~(\ref{eq:7})) and can be found by measuring it [Figure~\ref{fig1}]. It is important that each revival have a maximum in its own time interval. The $n$-th revival has a maximum in the interval from $(n-1) T_R$ to $n T_R$, and in this time interval the oscillator amplitude differs very little from the revival amplitude [Figure~\ref{fig1}]. Thus, by measuring the oscillator amplitude at different time intervals, we can determine the time dependence of the amplitudes of different revivals.

\begin{figure}[htbp]
\centering\includegraphics[width=1.0\linewidth]{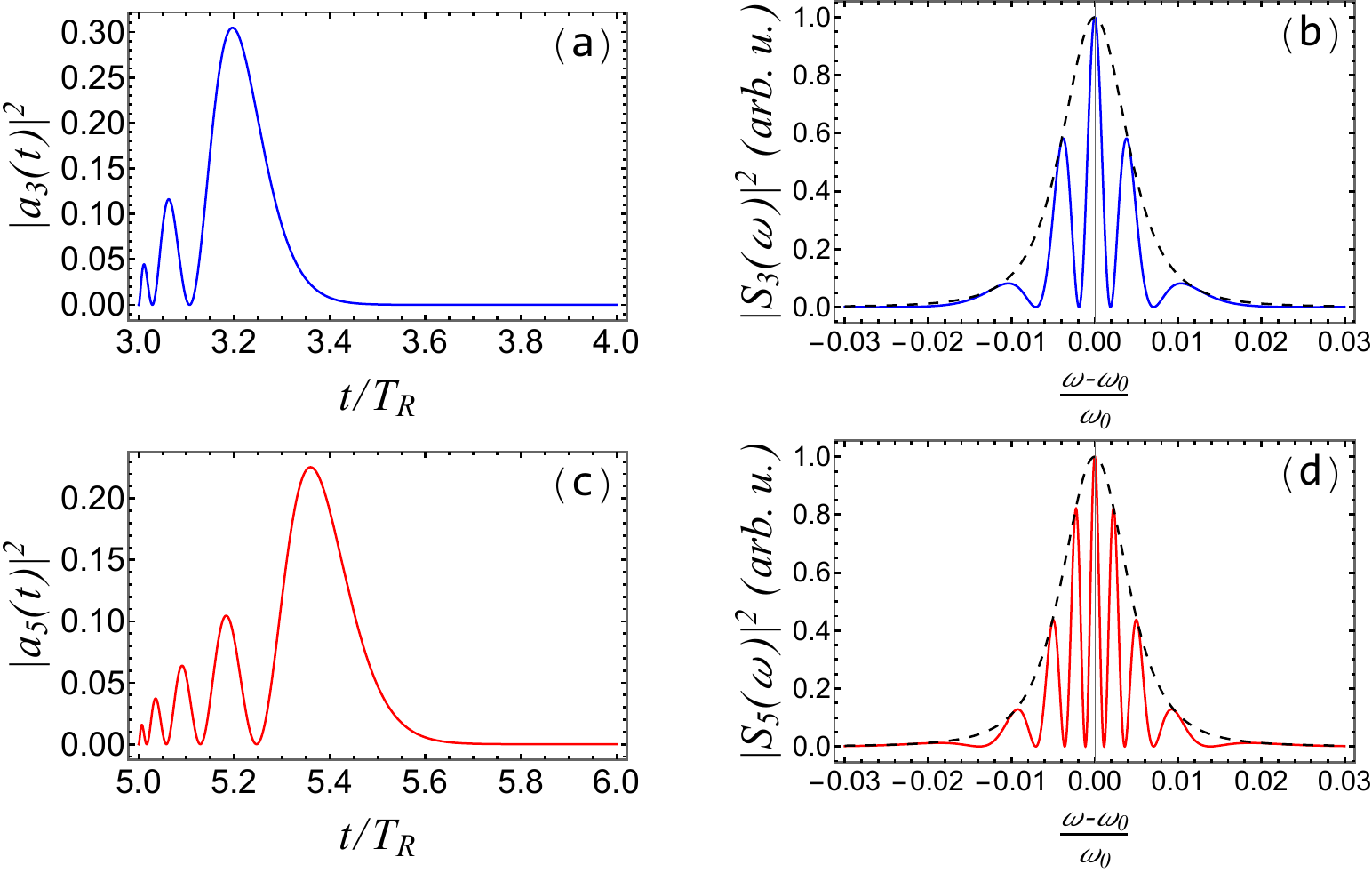}
\caption{Dynamics of the third $|a_3 (t)|^2$ (a) and fifth $|a_5 (t)|^2$ (c) revivals and their frequency spectra $|S_3(\omega)|^2$ (b) and $|S_5 (\omega)|^2$ (d). The black dashed line corresponds to spectrum of the initial evolution $a_0(t)$, namely, the Lorentz distribution $S_0 (\omega)=\frac{\gamma}{\pi}\frac{1}{\omega^2 + \gamma^2}$ . Here $\delta\omega=0.002 \omega_0$, $\gamma \approx 0.0071 \omega_0$, and $\omega_0=1$.}
\label{fig2}
\end{figure}

The amplitudes of different revivals have different time dependencies (see Eq.~(\ref{eq:9})). Measuring the time dependence for the $n$-th revival allows us to detect deviations from the exponential law and identify the exceptional point of the corresponding order. The deviations from the exponential law in the behavior of revival amplitudes can be clearly detected in their frequency spectra [Figure~\ref{fig2}]:
\begin{equation}
S_n (\omega)=Re\left(\frac{\gamma}{\pi}\frac{(\omega-i \gamma)^{n-1}}{(\omega+i\gamma)^{n+1}}\right)
\label{eq:12}
\end{equation}
These deviations lead to the appearance of additional peaks, the number of which is proportional to the order of the exceptional point. At the same time, the peak widths are inversely proportional to the order of the EP, which coincides with the revival numbers. Note that the frequency spectra of the different revivals can be calculated by taking the Fourier transform of the oscillator amplitude over the corresponding time intervals. Thus, it is possible to observe exceptional points of arbitrary order when considering the dynamics of the system at different time intervals.

In the experiment, the system under consideration can be implemented on the basis of a qubit interacting with a waveguide bounded on both sides by mirrors \cite{ferreira2021collapse} or a single-mode cavity interacting with a ring resonator \cite{hodaei2017enhanced, kullig2023higher, hodaei2014parity}. In such systems, the qubit (single-mode cavity) plays the role of the oscillator, and the waveguide (the ring resonator) acts as the reservoir. Such systems can serve as a convenient basis for observing non-Markovian effects, including those leading to the emergence of revivals \cite{ferreira2021collapse}.

\textit{Conclusion.} In conclusion, we consider the system of the harmonic oscillator interacting with the finite-sized reservoir. Non-Markovian effects caused by the finite size of the reservoir take place in such a system. We demonstrate that the non-Markovian effects lead to non-exponential dependence of the oscillator's amplitude on time. This behavior is typical for the one in the exceptional points. We demonstrate that the revivals taking place in the oscillator's dynamics can be described by the non-Hermitian equations with the exceptional point. The system dynamics are divided into intervals corresponding to different revivals, during each of which the ones are described by the product of the exponential and a power function. The higher the number of the revival, the higher the maximum degree of the power function and the order of the exceptional point in the differential equations describing the system dynamics over a given time interval. The deviation of the dynamics from the exponential law leads to modification of the spectrum, in particular, to the appearance of numerous additional peaks, the number of which is proportional to the order of the exceptional point. The peak widths are inversely proportional to the order of the EP. Thus, we demonstrate that by choosing the observation time, it is possible to observe the behavior that corresponds to the exceptional points of arbitrary order. We show that this behavior is a consequence of the non-Markovian effects.

\section*{Acknowledgments}
The study was financially supported by a Grant from Russian Science Foundation (Project No. 23-42-10010, https://rscf.ru/en/project/23-42-10010/). T.T.S., E.S.A, and A.A.Z. thank foundation for the advancement of theoretical physics and mathematics “Basis”.

\bibliography{apssamp}

\clearpage
\newpage

\onecolumngrid
\section*{Supplementary materials for "Producing exceptional points of arbitrary orders using non-Markovian effects"}
\subsection{Derivation of the non-Hermitian equation in Born-Markovian approximation}
To derive the non-Hermitian equation determining the dynamics of oscillator's amplitude, we first formally integrate the equation~(\ref{eq:3}):
\begin{equation}
{b_j (t)} = {b_j}(0){e^{ - i{\omega _j}t}} - ig\int\limits_0^t {d\tau {a}(\tau ){e^{ - i{\omega _j}(t - \tau )}}}
\label{eq:1A0}
\end{equation}
Then after substituting the Eq.~(\ref{eq:1A0}) into the Eq.~(\ref{eq:2}), we have:
\begin{equation}
{\dot a} =  - i{\omega _0}{a} - \sum\limits_j {{g^2}\int\limits_0^t {d\tau {a}(\tau ){e^{ - i{\omega _j}(t - \tau )}}} }
\label{eq:2A0}
\end{equation}
Here we use the assumption that ${b_j}(0) = 0$.

To proceed further, we should consider the reservoir to be infinite (specifically, here $\delta\omega \rightarrow 0$). Then we can use the Born-Markov approximation \cite{carmichael1999statistical,gardiner2004quantum} and calculate the integral in Eq. (\ref{eq:2A0}) using the Sokhotskii-Plemelj theorem \cite{plemelj1964}. Then we obtain the following non-Hermitian equation for the oscillator's amplitude:
\begin{equation}
{\dot a} =  - i{\omega _0}{a} - \gamma a
\label{eq:3A0}
\end{equation}
where $\gamma=\sum\limits_{j=-N/2}^{N/2} {\pi g^2 \delta(\omega_j-\omega_0)} =\pi g^2/\delta\omega$ is an effective decay rate. Note that this approximation works only in limit of infinite density of modes ($\delta\omega \rightarrow 0$). So in the case of finite $\delta\omega$, this approximation is applicable only at times $t<2\pi/\delta\omega$.

\subsection{Derivation of the expressions for the eigenstates of the system}
We consider the following systems of equations for averages of the system's operators:

\begin{equation}
\frac{{d{a}}}{{dt}} =  - i{\omega _0}{a} -i\sum\limits_{j =-N/2}^{N/2} {g{b_j}}
\label{eq:1A}
\end{equation}

\begin{equation}
\frac{{d{b_j}}}{{dt}} =  - i{\omega_j}{b_j} - ig{a}
\label{eq:2A}
\end{equation}
where $a(t) = \langle \hat a (t) \rangle$ and $b_j(t)=\langle \hat b_j (t) \rangle$.

Considering the initial state as $a(0)=1$ and $b_j(0)=0$ for all $j$, we can write the solution of the system (\ref{eq:1A})-(\ref{eq:2A}) as follows:
\begin{equation}
a(t)=\sum\limits_k {h_{ak}^2 e^{-i \tilde\omega_k t}}
\label{eq:3A}
\end{equation}

\begin{equation}
b_j(t)=\sum\limits_k {h_{jk}h_{ak} e^{-i\tilde\omega_k t}}
\label{eq:4A}
\end{equation}
where $h_{ak}$ and $h_{jk}$ are components of the eigenvector $\boldsymbol{h}_k = \{h_{ak},\ h_{(-N/2)k},...,\ h_{jk}, ...,\ h_{(N/2)k}\}^T$. $\tilde\omega_k$ is an corresponding eigenfrequency. Since the system is Hermitian, eigenfrequencies are real and all eigenvalues are mutually orthogonal. The equations for eigenstates have the following form:
\begin{equation}
(\omega_0 - \tilde\omega_k)h_{ak} + \sum\limits_{j=-N/2}^{N/2} {g h_{jk}}=0
\label{eq:5A}
\end{equation}
\begin{equation}
gh_{ak} + (\omega_j - \tilde\omega_k)h_{jk} =0
\label{eq:6A}
\end{equation}

From eq. (\ref{eq:6A}) we have $h_{jk}=\frac{g}{\tilde\omega_k - \omega_j} h_{ak}$. Substituting this relation into the equation (\ref{eq:5A}) we obtain the following equation, which determines the eigenfrequencies of the system:
\begin{equation}
\tilde\omega_k - \omega_0 = \sum\limits_{j=-N/2}^{N/2} {\frac{g^2}{\tilde\omega_k - \omega_j}}
\label{eq:7A}
\end{equation}

To calculate the eigenfrequencies we represent them as $\tilde\omega_k = \omega_k + \alpha_k = \omega_0 +k\delta\omega + \alpha_k$. Note that according to Sturmian separation theorem \cite{bellman1997introduction} $|\alpha_k| \le \delta\omega$. We consider the case $k \ge 0$ (the case $k \le 0$ is similar). Then the equation (\ref{eq:7A}) turns into the following expression:
\begin{equation}
\begin{gathered}
\alpha_k + k\delta\omega = \sum\limits_{j=-N/2}^{N/2} {\frac{g^2}{\alpha_k + (k-j)\delta\omega}}=\frac{g^2}{\alpha_k} + \sum\limits_{j \ne k} {\frac{g^2}{\alpha_k + (k-j)\delta\omega}}
\end{gathered}
\label{eq:8A}
\end{equation}
which can be written as:
\begin{equation}
\begin{gathered}
\alpha_k + k\delta\omega = \frac{g^2}{\alpha_k} + \sum\limits_{\substack {j'=-(\frac{N}{2}-k) \\ j' \ne 0}}^{\frac{N}{2}-k} {\frac{g^2}{\alpha_k - j'\delta\omega}} + \sum\limits_{j'=-(\frac{N}{2}-k)-1}^{-(\frac{N}{2}+k)} {\frac{g^2}{\alpha_k - j'\delta\omega}}
\end{gathered}
\label{eq:9A}
\end{equation}

Using the Taylor expansion, for the first sum in the right part of the equation (\ref{eq:9A}) we have:
\begin{equation}
\begin{gathered}
\sum\limits_{\substack {j'=-(\frac{N}{2}-k) \\ j' \ne 0}}^{\frac{N}{2}-k} {\frac{g^2}{\alpha_k - j'\delta\omega}} = \sum\limits_{j' \ne 0} \frac{-g^2}{j'\delta\omega} \sum\limits_{l=0}^{\infty} {\bigg(\frac{\alpha_k}{\delta\omega}\bigg)^l \frac{1}{{j'}^l}}
\end{gathered}
\label{eq:10A}
\end{equation}
Then using the fact that $\sum\limits_{\substack {j'=-(\frac{N}{2}-k) \\ j' \ne 0}}^{\frac{N}{2}-k} {j'}^{-l} =0$ for any odd $l$, we can obtain:
\begin{equation}
\begin{gathered}
\sum\limits_{j' \ne 0} \frac{-g^2}{j'\delta\omega} \sum\limits_{l=0}^{\infty} {\bigg(\frac{\alpha_k}{\delta\omega}\bigg)^l \frac{1}{{j'}^l}} = \frac{-2g^2}{\delta\omega} \sum\limits_{l=1}^{\infty} \sum\limits_{j'=1}^{\frac{N}{2}-k} {\bigg( \frac{\alpha_k}{\delta\omega} \bigg)^{2l-1} \frac{1}{{j'}^{2l}}}= \frac{-2g^2}{\alpha_k} \sum\limits_{l=1}^{\infty} {\bigg( \frac{\alpha_k}{\delta\omega}\bigg)^{2l} \sum\limits_{j'=1}^{\frac{N}{2}-k} \frac{1}{{j'}^{2l}}}=\\ \frac{-2g^2}{\alpha_k} \sum\limits_{l=1}^{\infty} {\bigg( \frac{\alpha_k}{\delta\omega}\bigg)^{2l} \bigg( \sum\limits_{j'=1}^{\infty} {\frac{1}{{j'}^{2l}}} - \sum\limits_{j'=\frac{N}{2} - k +1}^{\infty} {\frac{1}{{j'}^{2l}}} \bigg)}=\frac{-2g^2}{\alpha_k} \sum\limits_{l=1}^{\infty} {\bigg( \frac{\alpha_k}{\delta\omega}\bigg)^{2l} \bigg( \zeta(2l) - \zeta(2l,\frac{N}{2}-k +1) \bigg)}
\end{gathered}
\label{eq:11A}
\end{equation}
here we use the definitions of the Hurwitz $\zeta (s,a)=\sum\limits_{k=0}^{\infty} \frac{1}{(k+a)^s}$ and Riemann $\zeta(s)=\zeta(s,1) = \sum\limits_{k=1}^{\infty} \frac{1}{k^s}$ zeta functions.

The second sum in the right part of the equation (\ref{eq:9A}) can be rewritten in a similar way:
\begin{equation}
\begin{gathered}
\sum\limits_{j'=-(\frac{N}{2}-k)-1}^{-(\frac{N}{2}+k)} {\frac{g^2}{\alpha_k - j'\delta\omega}} =  \sum\limits_{l=0}^{\infty}\frac{-g^2}{\alpha_k} \frac{(-1)^{l+1} \alpha_k^{l+1}}{\delta\omega^{l+1}} \bigg( \zeta(l+1,\frac{N}{2}-k+1) - \zeta(l+1,\frac{N}{2}+k+1) \bigg)
\end{gathered}
\label{eq:12A}
\end{equation}

After substituting the expressions (\ref{eq:11A}) and (\ref{eq:12A}) into the equation (\ref{eq:9A}) we have:
\begin{equation}
\begin{gathered}
\alpha_k +k\delta\omega = \frac{g^2}{\alpha_k} -\frac{2g^2}{\alpha_k} \sum\limits_{l=1}^{\infty}{\bigg( \frac{\alpha_k}{\delta\omega} \bigg)^{2l} \zeta(2l)} +  \frac{2g^2}{\alpha_k} \sum\limits_{l=1}^{\infty}{\bigg( \frac{\alpha_k}{\delta\omega} \bigg)^{2l}} \zeta(2l,\frac{N}{2}-k+1) - \\ \frac{g^2}{\alpha_k}\sum\limits_{l=0}^{\infty} \frac{(-1)^{l+1} \alpha_k^{l+1}}{\delta\omega^{l+1}} \bigg( \zeta(l+1,\frac{N}{2}-k+1) - \zeta(l+1,\frac{N}{2}+k+1) \bigg)
\end{gathered}
\label{eq:13A}
\end{equation}

In order to turn this equation into a more convenient form, we should recall the following two relations:
\begin{equation}
\sum\limits_{k=1}^{\infty} {t^{2k}\zeta(2k)} = \frac{1}{2} (1-\pi t \cot(\pi t))
\label{eq:14A}
\end{equation}
and
\begin{equation}
\psi^{(n)}(x)= (-1)^{n+1} n!\ \zeta(n+1,x)
\label{eq:15A}
\end{equation}
where $\psi (x)$ is the digamma function.

The first sum in the right part of the equation (\ref{eq:13A}) can be simplified using the relation (\ref{eq:14A}):
\begin{equation}
\frac{2g^2}{\alpha_k} \sum\limits_{l=1}^{\infty}{\bigg( \frac{\alpha_k}{\delta\omega} \bigg)^{2l} \zeta(2l)} = \frac{g^2}{\alpha_k} - \frac{\pi g^2}{\delta\omega} \cot(\frac{\pi \alpha_k}{\delta\omega})
\label{eq:16A}
\end{equation}

For the second sum in Eq. (\ref{eq:13A}) we use the relation (\ref{eq:15A}) and obtain:
\begin{equation}
\begin{gathered}
\frac{2g^2}{\alpha_k} \sum\limits_{l=1}^{\infty}{\bigg( \frac{\alpha_k}{\delta\omega} \bigg)^{2l}} \zeta(2l,\frac{N}{2}-k+1) = \frac{2g^2}{\delta\omega} \sum\limits_{l=1}^{\infty} {\bigg( \frac{\alpha_k}{\delta\omega} \bigg)^{2l-1} (-1)^{2l} \zeta(2l,\frac{N}{2}-k+1)} = \\ \frac{2g^2}{\delta\omega} \sum\limits_{l=1}^{\infty} {\frac{1}{(2l-1)!} \psi^{(2l-1)}(\frac{N}{2}-k+1) \bigg(\frac{\alpha_k}{\delta\omega} \bigg)^{2l-1}}=\\ \frac{g^2}{\delta\omega}\bigg[\sum\limits_{l=0}^{\infty} {\frac{1}{l!} \psi^{(l)}(\frac{N}{2}-k+1) \bigg(\frac{\alpha_k}{\delta\omega} \bigg)^{l}} -  \sum\limits_{l=0}^{\infty} {\frac{1}{l!} \psi^{(l)}(\frac{N}{2}-k+1) \bigg(-\frac{\alpha_k}{\delta\omega} \bigg)^{l}}  \bigg] =\\ \frac{g^2}{\delta\omega} \bigg[ \psi(\frac{N}{2} - k +1 + \frac{\alpha_k}{\delta\omega}) - \psi(\frac{N}{2} - k +1 - \frac{\alpha_k}{\delta\omega}) \bigg]
\end{gathered}
\label{eq:17A}
\end{equation}
here we used the fact that $\sum\limits_{l=0}^{\infty}{\frac{1}{l!} \psi^{(l)} (x_0) (x-x_0)^l} = \psi(x)$ is the Taylor expansion of the digamma function.

And for the third sum, similarly, we have:
\begin{equation}
\begin{gathered}
\frac{g^2}{\alpha_k}\sum\limits_{l=0}^{\infty} \frac{(-1)^{l+1} \alpha_k^{l+1}}{\delta\omega^{l+1}}  \bigg( \zeta(l+1,\frac{N}{2}-k+1) - \zeta(l+1,\frac{N}{2}+k+1) \bigg) =\\ \frac{g^2}{\delta\omega} \bigg[ \psi(\frac{N}{2} - k +1 + \frac{\alpha_k}{\delta\omega}) - \psi(\frac{N}{2} + k +1 + \frac{\alpha_k}{\delta\omega}) \bigg]
\end{gathered}
\label{eq:18A}
\end{equation}

After substituting the expressions (\ref{eq:16A}), (\ref{eq:17A}) and (\ref{eq:18A}) into the Eq. (\ref{eq:13A}) we obtain the following equation:
\begin{equation}
\alpha_k +k\delta\omega = \gamma \cot(\frac{\pi \alpha_k}{\delta \omega}) + \frac{\gamma}{\pi} \bigg[ \psi(\frac{N}{2} + k +1 + \frac{\alpha_k}{\delta\omega}) - \psi(\frac{N}{2} - k +1 - \frac{\alpha_k}{\delta\omega}) \bigg]
\label{eq:19A}
\end{equation}
where $\gamma=\pi g^2 /\delta\omega$. Note that $\gamma$ here coincides with the effective decay rate $\gamma= \sum\limits_{j=-N/2}^{N/2} {\pi g^2 \delta(\omega_j - \omega_0)} = \pi g^2 /\delta\omega$, obtained through elimination of reservoir's degrees of freedom within the Born-Markov approximation.

This transcendental equation exactly determines the eigenfrequencies of the considered system. However, it is more convenient to consider the case of a large number of modes in the reservoir $N \rightarrow \infty$ (note that here we can fix the step between modes' frequencies $\delta\omega$, i.e. the case when frequencies span the entire range from $-\infty$ to $+\infty$). In this case the second term in the right part of the equation (\ref{eq:19A}) tends to zero. Then we can get a simplified equation:
\begin{equation}
\alpha_k +k\delta\omega = \gamma \cot(\frac{\pi \alpha_k}{\delta \omega})
\label{eq:20A}
\end{equation}
or, equivalently
\begin{equation}
\alpha_k = \frac{\delta\omega}{\pi}\arctan\bigg(\frac{\gamma}{\alpha_k + k\delta\omega}\bigg)
\label{eq:21A}
\end{equation}

The dynamics of the oscillator $a(t)$ is determined by eq. (\ref{eq:3A}) and by eigenvector's components $h_{ak}^2$. To calculate these components, we consider all eigenvectors to be normilized, i.e. $h_{ak}^2 + \sum\limits_j {h_{jk}^2} = 1$. After substituting the relation (\ref{eq:6A}) into this equation we have:
\begin{equation}
h_{ak}^2 = \frac{1}{1+\sum\limits_j {\frac{g^2}{(\tilde\omega_k - \omega_j)^2}}}
\label{eq:22A}
\end{equation}
From the equation (\ref{eq:20A}) we also have:
\begin{equation}
\begin{gathered}
\sum\limits_j {\frac{g^2}{(\tilde\omega_k - \omega_j)^2}} = -\frac{\partial}{\partial\alpha_k} \sum\limits_j {\frac{g^2}{\alpha_k +(k-j)\delta\omega}}=-\frac{\partial}{\partial\alpha_k} \gamma\cot(\frac{\pi \alpha_k}{\delta\omega})
\end{gathered}
\label{eq:23A}
\end{equation}
Then we substitute this expression into the eq. (\ref{eq:22A}) and obtain:
\begin{equation}
h_{ak}^2 = \frac{\tan^2 (\frac{\pi \alpha_k}{\delta\omega})}{(1+\pi\gamma/\delta\omega)\tan^2 (\frac{\pi \alpha_k}{\delta\omega}) + \pi\gamma/\delta\omega}
\label{eq:24A}
\end{equation}
Now we can use the eq. (\ref{eq:21A}) and obtain the following expression for the components $h_{ak}^2$:
\begin{equation}
\begin{gathered}
h_{ak}^2 = \frac{1}{\pi}\frac{\gamma\delta\omega}{(\alpha_k + k\delta\omega)^2 + \gamma^2 (1+\delta\omega/\pi\gamma)}= \frac{1}{\pi} \frac{\gamma\delta\omega}{(\tilde\omega_k - \omega_0)^2 +\Gamma^2}
\end{gathered}
\label{eq:25A}
\end{equation}
where $\Gamma^2 = \gamma^2 (1+\delta\omega/\pi \gamma)$. Thus, the components $h_{ak}^2$ have Lorentz-like distribution with a linewidth of $\Gamma$.

\subsection{Dynamics of the system. Collapses and revivals of the oscillations}
The dynamics of the oscillator's amplitude is determined by the Eq. (\ref{eq:3A}). Using the expression (\ref{eq:25A}) we can obtain the following equation (for simplicity, we switch to slow amplitudes, which is equivalent to the case of $\omega_0 =0$):
\begin{equation}
a(t)=\sum\limits_k {\frac{1}{\pi}\frac{\gamma\delta\omega}{\omega_k^2 + \Gamma^2} e^{-i \tilde\omega_k t}}
\label{eq:1B}
\end{equation}
As we stated before, for simplicity, we consider the number of modes to be infinite, so the summation in Eq. (\ref{eq:1B}) goes from $-\infty$ to $+\infty$. In order to calculate this sum we use the Poisson summation formula \cite{fleischhauer1993revivals}:
\begin{equation}
a(t)=\sum\limits_{n=-\infty}^{\infty}{\int\limits_{-\infty}^{\infty} dk \frac{1}{\pi}\frac{\gamma\delta\omega}{\tilde\omega^2(k) + \Gamma^2} e^{-i \tilde\omega(k)t} e^{i2\pi n k}}
\label{eq:2B}
\end{equation}

To calculate the presented integral it is more convenient to switch from the integration over the modes' number $dk$ to integration over frequencies $d\omega$. Hereinafter, we use the notation $\tilde\omega_k \rightarrow \omega$. By definition we have $\tilde\omega_k = \omega=k\delta\omega + \alpha_k$. Then, by considering the equation (\ref{eq:21A}), we have:
\begin{equation}
k=\frac{\omega}{\delta\omega} - \frac{1}{\pi} \arctan\bigg(\frac{\gamma}{\omega}\bigg)
\label{eq:3B}
\end{equation}
and
\begin{equation}
dk=\frac{d\omega}{\delta\omega} + \frac{\gamma d\omega}{\pi(\omega^2 + \gamma^2)}=\frac{\omega^2 +\Gamma^2}{\omega^2 + \gamma^2}\frac{d\omega}{\delta\omega}
\label{eq:4B}
\end{equation}

After substituting the expressions (\ref{eq:3B}) and (\ref{eq:4B}) into the equation (\ref{eq:2B}) we have:
\begin{equation}
a(t)=\sum\limits_{n=-\infty}^{\infty}{\int\limits_{-\infty}^{\infty} d\omega \frac{\gamma}{\pi}\frac{e^{-i\omega(t-\frac{2\pi n}{\delta\omega})}}{\omega^2 + \gamma^2} e^{-2in \arctan(\frac{\gamma}{\omega})}}
\label{eq:5B}
\end{equation}
Using the logarithmic representation of the arctangent function $\arctan(z)=\frac{i}{2} \ln(\frac{i+z}{i-z})$ we can obtain:
\begin{equation}
a(t)=\sum\limits_{n=-\infty}^{\infty}{\int\limits_{-\infty}^{\infty} d\omega \frac{\gamma}{\pi}\frac{e^{-i\omega(t-\frac{2\pi n}{\delta\omega})}}{\omega^2 + \gamma^2} \frac{(\omega-i\gamma)^n}{(\omega+i\gamma)^n}}
\label{eq:6B}
\end{equation}

The integrals
\begin{equation}
a_n(t-\frac{2\pi n}{\delta\omega})=\int\limits_{-\infty}^{\infty} d\omega \frac{\gamma}{\pi}\frac{e^{-i\omega(t-\frac{2\pi n}{\delta\omega})}}{\omega^2 + \gamma^2} \frac{(\omega-i\gamma)^n}{(\omega+i\gamma)^n}
\label{eq:7B}
\end{equation}
can be calculated using the residue theorem. From physical point of view we are interested only in positive time intervals. So in all calculations we use the fact that $t>0$. Thus, one can notice that for all $n<0$ the integrals (\ref{eq:7B}) equal to zero, since the integrand functions have no poles in the lower half of complex plane. The same situation for the case $n>0$ at times $t<\frac{2\pi n}{\delta\omega}$ (integrand functions have no poles in the upper half of complex plane). So, the integrals (\ref{eq:7B}) are non-zero only when $n \ge 0$ and $t>\frac{2\pi n}{\delta\omega}$.

Calculation of these integrals gives us the following result:
\begin{equation}
a(t)=e^{-\gamma t} + \sum\limits_{n=1}^{\infty} \theta(t-\frac{2\pi n}{\delta\omega})a_n (t-\frac{2\pi n}{\delta\omega})
\label{eq:8B}
\end{equation}
where $\theta(x)$ is the Heaviside step function and
\begin{equation}
a_n (t) = -\frac{2i\gamma}{n!}\frac{d^n}{d\omega^n}\bigg[ e^{-i\omega t} (\omega - i\gamma)^{n-1} \bigg] \bigg|_{\omega = -i\gamma}
\label{eq:9B}
\end{equation}

The moments of time $t=\frac{2\pi n}{\delta\omega}$ correspond to revivals of the energy in the dynamics of $a(t)$. Indeed, if we look for the first few terms in the sum (\ref{eq:8B}) we can get the following expression:
\begin{equation}
\begin{gathered}
a(t)=e^{-\gamma t} - \theta(t-\frac{2\pi}{\delta\omega})2\gamma \bigg(t-\frac{2\pi}{\delta\omega}\bigg) e^{-\gamma(t-\frac{2\pi}{\delta\omega})} +  \theta(t-\frac{4\pi}{\delta\omega})2\gamma^2 \bigg(t-\frac{4\pi}{\delta\omega}\bigg)  \bigg(\bigg(t-\frac{4\pi}{\delta\omega}\bigg)-\frac{1}{\gamma}\bigg)e^{-\gamma(t-4\pi/\delta\omega)} - ...
\end{gathered}
\label{eq:10B}
\end{equation}

From the Eq. (\ref{eq:9B}) we can notice that the functions $a_n(t)$ can also be determined by the following recurrence formula:
\begin{equation}
a_n (t) = \frac{n-1}{n}a_{n-1} (t) - \frac{2\gamma t}{n} \sum\limits_{k=0}^{n-1} {a_{k} (t)}
\label{eq:11B}
\end{equation}
where $n \ge 1$ and $a_0 (t)=e^{-\gamma t}$.

The recurrence relation (\ref{eq:11B}) can be rewritten in a following way:
\begin{equation}
a_{n+2} (t)= \frac{2(n+1)}{n+2}\bigg( 1-\frac{\gamma t}{n+1} \bigg)a_{n+1} (t) - \frac{n}{n+2}a_n (t) 
\label{eq:12B}
\end{equation}
This recurrence equation coincides with the one for the generalized Laguerre polynomials $L_n^{(-1)}(x)$. Considering that $a_0 (t)=e^{-\gamma t}$, we have:
\begin{equation}
a_{n} (t)= e^{-\gamma t} L_n^{(-1)} (2\gamma t)
\label{eq:13B}
\end{equation}
Thus, the solution for oscillator's amplitude is determined by the following expression:
\begin{equation}
a(t)= \sum\limits_{n=0}^{\infty} {\theta(t-\frac{2\pi n}{\delta\omega}) e^{-\gamma (t-\frac{2\pi n}{\delta\omega})} L_n^{(-1)} \bigg(2\gamma (t-\frac{2\pi n}{\delta\omega})\bigg)}
\label{eq:14B}
\end{equation}

From equation (\ref{eq:9B}) we can also obtain the following chain of equations for the amplitudes $a_n (t)$:
\begin{equation}
\dot a_n (t)=-\gamma a_n (t) - 2\gamma \sum\limits_{k=0}^{n-1} {a_k (t)}
\label{eq:15B}
\end{equation}
or, equivalently, in matrix form:
\begin{equation}
\frac{d}{{dt}}\left( {\begin{array}{*{20}{c}}
  {{a_0}} \\ 
  {{a_1}} \\ 
  {a_2} \\ 
  {\vdots } \\
  {a_n}
\end{array}} \right) = \left( {\begin{array}{*{20}{c}}
  { - \gamma}&{ 0 }&{ 0}&{ \dots}&{0} \\ 
  { - 2\gamma }&{ -\gamma}&{ 0 }&{ \dots}&{0} \\ 
  {-2\gamma }&{-2\gamma}&{-\gamma}&{\dots}&{0} \\ 
  {\vdots }&{\vdots}&{\vdots }&{\ddots} \\
  {-2\gamma }&{-2\gamma}&{-2\gamma}&{\dots}&{-\gamma}
\end{array}} \right)\left( {\begin{array}{*{20}{c}}
  {{a_0}} \\ 
  {{a_1}} \\ 
  {a_2 } \\ 
  {\vdots } \\
  {a_n}
\end{array}} \right)
\label{eq:15B}
\end{equation}
here the initial condition is $a_0 (0)=1$ and $a_k (0)=0$, $k>0$. Note that the equation for $a_0$ coincides with the equation (\ref{eq:3A0}) (considering $\omega_0 =0$).

\nocite{*}
% Produces the bibliography via BibTeX.
\end{document}